\documentclass[12pt]{article}
\usepackage{00}

\begin{document}

\title{Empirical Confirmation (and Refutation) \\
        of Presumptions on Software
}
\author{Yossi Gil
\thanks{Work done in part while author was with IBM Research--Haifa}\\
  {\emph{The Technion}---Israel Institute of Technology}\\
  \texttt{yogi@cs.technion.ac.il}
}

\author{Maayan Goldstein\\
  IBM Research---Haifa\\
  \texttt{maayang@il.ibm.com}
}
\author{Dany Moshkovich\\
  {IBM Research---Haifa}\\
  \texttt{mdany@il.ibm.com}
}

\maketitle

\begin{abstract}
Code metrics are easy to define, but not so easy
  to justify.
It is hard to prove that a metric is valid,
  i.e., that measured numerical values imply anything
  on the vaguely defined, yet crucial software properties
  such as complexity and maintainability.
This paper employs statistical analysis and tests to check some ``believable''
presumptions on the behavior of software and metrics measured for this software.
Among those are the reliability presumption
  implicit in the application of any code metric, and the presumption that the magnitude
  of change in a software artifact is correlated with changes to its version number.

Putting a suite of \metricsCount{} metrics to the trial,
  we confirm most of the presumptions.
Unexpectedly, we show that a substantial portion
  of the reliability of some metrics can be observed even
  in random changes to architecture.
Another surprising result is that Boolean-valued metrics tend to flip their values
  more often in minor software version increments than in major increments.

\end{abstract}

\paragraph{keywords}
metrics; reliability; software architecture

\renewcommand{\thefootnote}{\arabic{footnote}}

%\tableofcontents

\section{Introduction}
\label{Section:introduction}
\subsection{Metrics' Reliability}
Software metrics are considered~\cite{Lorenz1994OOSoftwareMetricsGuide} an important tool
     of software engineering.
However, just as with any other kind of measurement, software metrics are subject to
\emph{reliability} and \emph{validity} concerns.
The main cast of doubt in using metrics
    is that of validity;
``What do these numbers mean?'',
  ``how do they reflect on quality?'', ``complexity?'',
  are typical questions that one would ask when
  bombarded with a list of metric values.

In contrast, the issue of \emph{reliability},
  even if occasionally mentioned~\cite{rajaraman92,Meyer1998,KitchenhamS97,Garcia2004},
  was largely dismissed~\cite{ChidamberK94,Mauricio2008StateOfMetrics,%
      Lajios2009MetricsSuites,Lorenz1994OOSoftwareMetricsGuide,
  Henderson1996ObjectOrientedMetrics,Fenton1996SoftwareMetrics}.
Reliability concern is believed to be associated more with psychological
  and biological measurements, such as performance in an IQ test or human evaluation of
  an X-ray scan---in which measurement errors are inherent.
However, people compute software metrics  without
    worrying about reliability---after all,
    how can the computation of the number of lines of code
    yield an inaccurate result?

One of the presumptions this work examines is the hidden presumption of absolute
    reliability in measurement and ensuing evaluation of software.
We maintain that such measurement, per se, is accurate,
  and of course do not argue e.g., that quantum phenomena
  may inject errors into the underlying computations;
  nor do we care here about defects in the implementation of
  metric algorithms.
However we suggest that in the face
    of software changes, the method  of ``instantaneous capture'' applied
   in the computation of metrics,  should be scrutinized.
The motivation should be clear: long gone are the days that
	software was produced, frozen,
  and then used without any subsequent changes.
With the increasing shift to agile development
  process~\cite{Goth2009Agile},
  changes to software are becoming even more frequent,
  and hence metric values have shorter lives' spans.

Say that a certain class's \emph{Depth in Inheritance Tree}
  is~3 in version 20.0-b11 of a certain software artifact,
  then,  some natural questions to ask are whether
  this value is good or bad, how it reflects on maintainability,
  whether it can be used to predict correctness, etc.
All these questions belong in the validity domain.
One can also ask whether there is a trend of increasing
  or decreasing of the \emph{Depth in Inheritance Tree}
  metric,  should this trend be encouraged or discouraged,
  whether the measured value is reflective of the entire artifact,
  typical of artifacts of this kind, etc.
However, there is a more fundamental reliability
    question to ask:
is this measurement stable with respect to natural
	evolution of the software?
In other words, one may ask what is the likelihood of finding
	 a different value in  version (say) 20.0-b11a of the said
    artifact.
There is little point in making any conclusions regarding
  any measurement if this measurement is subject to random
  fluctuations during software evolution.

  This observation brought us to investigating additional presumptions, that relate to changes
  in software size, correlation between changes in software versions and magnitude and nature of the changes in the code of the software artifacts.

Our study employed \metricsCount{} different code metrics, selected
  from several independent collections of metrics.
We organized these metrics in a taxonomy whose
  main groups are: \emph{marker} (or Boolean) metrics, \emph{local numerical} metrics and
  \emph{global numerical} topological metrics.
  To understand better what we mean by
    ``topological'' metrics,
    recall that programs can be readily
  represented as a directed graph of classes,
  packages, or other modules,
    which can then be subjected to graph theoretical algorithms~\cite{Tarjan72,Aho1974,Sedgewick1983}.

 To check the different presumptions, we used a large corpus of software versions, and applied
  the same set of metrics to each version.
We then asked whether the results are reliable, i.e., whether the values obtained
  in a certain version are predictive of the values
  in the subsequent version. We further investigated how changes in version size and number are correlated with
  the metrics. Finally, we examined how some presumptions change for different groups of metrics.

An intriguing finding of this work
    is that a substantial portion
  of the reliability of the global metrics can be observed even
  if random perturbations are applied to the architecture.
This means, in a sense, that these metrics do not capture
    an inherent architectural property of the software.

Another interesting result is that marker metrics tend to change less in major version increments and more
  in minor version changes. This may mean that major version releases are more stable and carefully organized than the minor ones.

\subsection{Platitudes}
The ``reliability'' presumption, which we were
    able to partially confirm, is just
    one of many hidden presumptions, or \emph{platitudes},
    as we shall interchangeably refer to these henceforth, regarding
    software evolution.
On course of our study, we were able
    to examine several of these, refuting some,
        and confirming the others.

The list of these platitudes is as follows:
first, everyone knows that
  \platitude{size-variety}{software comes in many different sizes}
There is an almost universal agreement on a Dewey like version numbering scheme,
  and people tend to believe that
  \platitude{major-changes-large}{changes to major version number are
      correlated with the magnitude of the software change}
It is likewise common knowledge that
  \platitude{evolutionary-revolutionary-spectrum}{software changes fall
  in a continuous range between the evolutionary
  end, at which most meaningful properties are preserved, and the
  antipodal revolutionary end}

The concepts of revolutionary and evolutionary changes may seem amorphous. However,
one may think of restructuring an existing software system to fit a model--view--controller
pattern~\cite{Gamma:1995:DPE} as
a revolutionary change, and of adding new encryption method to a banking system as an evolutionary change.

Further, it is plausible to assume that
  \platitude{mostly-evolutionary}{most releases of new software
    versions are evolutionary}
  and that \platitude{revolutionary-changes-in-major-versions}{revolutionary changes tend to coincide with changes to major version number}
We may also subscribe to beliefs regarding the kind of changes.
One would tend to think  that
\platitude{preservation-of-style}{additions to an existing software
  body tend to follow existing style; more so with evolutionary changes}
Also, it is believable that \platitude{locality-of-change}{even large changes
  to software tend to leave substantial isolated portions of the code unchanged}
And, of course, the tacit reliability presumption that we begun with is:
  \platitude{metrics-reliability}{metrics are reliable}.

Our results confirmed most of the presumptions. (For example, we found that
reliability of \texttt{final} or \texttt{abstract} is typically close to $100\%$.)
But, a number of very ``believable'' presumptions, including \Pla{locality-of-change} were refuted.

\subsection{Contributions}
The main contributions of this paper are:
\begin{enumerate}
\item Raising the somewhat less visited issue of \emph{software metrics reliability}.
\item An introduction of a taxonomy of code metrics.
\item The discovery of similarity in many of the properties of metrics in each group.
\item A systematic application of statistical methods to confirm (or refute) presumptions on software.
\item The revelation that local metrics are highly reliable.
\item The discovery that although global metrics tend to be reliable,
      much of this reliability is due to the limited scope of changes.
\item The discovery of the surprising fact that local metrics tend to change
          more often in minor version changes.
\item The revelation that local metrics are 99\% reliable.
\item The discovery of the link between the ranking imposed by numerical global metrics
      and the topological architecture, i.e., software properties which can be inferred by
  examining the structure of the software graph,
  but without using any semantical information.

\item The discovery that although global types tend to be reliable,
      much of this reliability is due to the limited scope of changes.
\end{enumerate}

\paragraph{Outline.}

The remainder of this article is organized as follows.
The data corpus and the way it was selected are described
  in \Sec{corpus}. This section also discusses the presumption of \Pla{size-variety}.
  \Sec{size} then analyzes the size changes
    of software artifacts present in the corpus,
    examining presumptions \Pla{locality-of-change}, \Pla{evolutionary-revolutionary-spectrum},
  \Pla{mostly-evolutionary} and \Pla{major-changes-large}.
This section takes an intermission to remind the reader
  of Kendall's tau correlation coefficient and its use as an indicator
  of similarity between rankings.
We will use this coefficient later also in the
    analysis of numerical metrics.

\Sec{metrics} presents
  our metrics suite and the way it was selected, and our metrics taxonomy.
In \Sec{marker} we study the reliability of
  the metrics that yield Boolean values, discussing the
	\Pla{mostly-evolutionary},
  \Pla{evolutionary-revolutionary-spectrum} and
	\Pla{preservation-of-style} presumptions.
These three presumptions are revisited in \Sec{results} which
  presents reliability results of numerical metrics.
The analysis that
    shows that at least part of the reliability
    of global numerical metrics cannot
    be attributed to inherent ``software architecture''
    is presented \Sec{mutation}.

Related work is the subject of Section~\ref{Section:related},
    while Section~\ref{Section:conclusions} concludes and
  suggests directions for further research.

\section{Software Corpus}
\label{Section:corpus}
\subsection{Artifacts}
The software corpus used in our experiments comprised~$\BatchSizeN$ software \emph{artifacts}, all drawn from the
  \emph{Qualitas Corpus}\footnote{See the \emph{Qualitas Research Group, Qualitas Corpus} \texttt{http://www.cs.\-auck\-land.\-ac.\-nz\-/\-\textasciitilde{}ewan/\-cor\-pus}},
  a colossal collection of \textsc{Java} software that is being used extensively
  in many empirical software engineering studies%
      \footnote{See \texttt{http://www.cs.auckland.ac.nz/\textasciitilde{}ewan/cor\-pus/publications.html} for a partial list.}.

These artifacts included:
  the \textsc{Java} compiler, \texttt{javac},
  \texttt{ant} (\textsc{Java}'s equivalent of \texttt{make}),
  and  \texttt{junit} (the \textsc{Java} unit testing library),
 \ignore{ \texttt{azureus} (a bit torrent client),}
   Eclipse's JDT core,
  search,
  and SWT, \texttt{FreeCol} (a simulation game),
  \texttt{Antlr} (a framework for constructing compilers, interpreters, etc.)
  \texttt{hibernate} (a persistence framework),
  \texttt{holds} (a relational database engine),
  \texttt{jgraph} (a graph drawing package),
  \texttt{log4j} (the logging component of Apache),
  \texttt{struts} (the Apache framework for the creation of web applications),
  \texttt{weak} (data mining and machine learning software),
  \texttt{argouml} (an UML diagramming application),
  \texttt{hsqldb} (hyper SQL database engine),
  \texttt{jhotdraw} (java GUI framework for technical and structured graphics),
  \texttt{jung} (framework for modeling, analysis and visualization of graphs), and
  \texttt{proguard} (java shrinker, optimizer, obfuscator and preverifier).

\subsection{Versions}
For each artifact, we analyzed a number of \emph{versions} from the corpus.
We harvested all available versions of each of the artifacts, omitting
  only three versions in which global renaming made it difficult to automatically
  trace classes of previous versions.
In total, our corpus comprised~\BatchSizeTotal{} versions.

The essential size characteristics of the corpus are summarized in \Tab{corpus}.
The corpus totaled some 78 thousands types,
  organized in 5,500 packages.
In agreement with \Pla{size-variety}, the number of types in the versions selected in the corpus spans
  two orders of magnitude (42 through 6,444),
  with a median and average of a few hundreds of types.
A similar variety is observed in the number of packages.

\begin{table}[!ht]
\centering
%\scriptsize
\renewcommand\arraystretch{1.3}
\renewcommand\pad{\hspace{0pt}}
\begin{tabular}{LCCCII}
%VLCCII
Size Metric	& \multicolumn{1}{H}{{Mean}}	& \multicolumn{1}{H}{{Median}}	& \multicolumn{1}{H}{{Min}}	& \multicolumn{1}{H}{{Max}}	& \multicolumn{1}{H}{{Total}}	\\
\hline % Header end	\\
Types	& $822{\pm 1,125}$	& $420{\pm 285}$	& $42$	& $6,444$	& $78,099$	\\
Packages	& $58{\pm 98}$	& $23{\pm 13}$	& $3$	& $469$	& $5,500$	\\
Edges	& $3,767{\pm 4,910}$	\mbox{ }& $2,069{\pm 1,437}$	& $77$	& $27,764$	& $357,897$	\\
\hline % Footer	\\

\end{tabular}
\caption{Size characteristics of the software corpus.}
\label{Table:corpus}
\end{table}

Each software version was modeled as a \emph{directed graph},
  in which types serve as nodes, and edges lead from a type to all types which
  it uses directly, i.e., inheriting from it, declaring a variable of it, invoking one of its methods, etc.
Edges leading to outside the artifact, e.g., the edge that leads from almost every \textsc{Java} class to
  \texttt{java.lang.Object}, were ignored.
The number of edges thus found is shown in the last row of the table.
Not surprisingly \Pla{size-variety}, we see a two orders of magnitude variety
  in the number of edges as well.

\Tab{corpus}
  introduces a~$\pm$ notation that embellishes the mean with
  the standard deviation, e.g., the \emph{mean} number of types is~$822$ (averaged over all~$\BatchSizeTotal$ software versions),
  while the standard deviation is~$1,125$.
Similarly, the median is embellished with the \emph{median absolute deviation} (M.A.D.),
  defined as the median of the absolute deviations from the median
  of the distribution.

The large standard deviation and the wide range of values are not surprising---software varies greatly in size.
For this reason, we prefer the median and the M.A.D.\ as a pair of summarizing statistics over the
  mean and standard deviation.
Admittedly, the median and the M.A.D are less efficient statistical measures
  than the mean and the standard deviation, but they are
  robust to outliers,  which are unavoidable with this great variety.

\subsection{Pairs}
Our study of software change was carried by organizing
  the~\BatchSizeTotal{} versions in the corpus in an ensemble of~\BatchConsecutivepairsTotal{}  \emph{pairs}
  of subsequent versions of the same artifact.

Some statistics of software growth and the extent of preservation in the pairs of the
  corpus are shown in \Tab{evolution}.

\begin{table}[!htb]
\centering
\renewcommand\arraystretch{1.1}
\renewcommand\pad{\hspace{3pt}}
\small
%\scriptsize
\begin{tabular}{/L!!II}
Metric	& \multicolumn{1}{H}{{Mean\mbox{ }(\%)}}	& \multicolumn{1}{H}{{Median\mbox{ }(\%)}}	& \multicolumn{1}{H}{{Min\mbox{ }(\%)}}	& \multicolumn{1}{H}{{Max\mbox{ }(\%)}}	\\
\hline % Header end	\\
Types	& 137!{\pm 64}	& 112!{\pm 11}	& 100	& 517	\\
Edges	& 146!{\pm 92}	& 118!{\pm 17}	& 90	& 712	\\
\hline %break	\\
Remaining Types	& 91!{\pm 16}	& 97!{\pm 3}	& 19	& 100	\\
Continuing Types	& 75!{\pm 23}	& 79!{\pm 17}	& 11	& 100	\\
Remaining Edges	& 86!{\pm 18}	& 94!{\pm 6}	& 23	& 100	\\
Continuing Edges	& 71!{\pm 26}	& 76!{\pm 20}	& 6	& 100	\\
\hline %break	\\
Unchanged Types (outgoing)	& 17!{\pm 7}	& 17!{\pm 5}	& 1	& 36	\\
Unchanged Types (incoming)	& 16!{\pm 7}	& 17!{\pm 4}	& 1	& 36	\\
Unchanged Types (both)	& 16!{\pm 7}	& 17!{\pm 5}	& 1	& 36	\\
\hline % Footer	\\

\end{tabular}
\caption{Growth and changes in consecutive artifact versions.}
\label{Table:evolution}
\end{table}

Table~\ref{Table:evolution} should help us appreciate the magnitude of changes
  and the extent of preservation in the version pairs used in our corpus.
On average, the number of types increased by~$37\%$ and the number of edges by~$46\%$.
Again, we observe a wide spectrum of changes, e.g., in one of the pairs, the number of edges
  increased by a factor of~$7$. There were even cases in which the number of edges decreased, probably thanks to code refactoring which reduced coupling between types.

  This great variety can be interpreted as a supportive indication of \Pla{evolutionary-revolutionary-spectrum}. Furthermore, the fact that in the first two lines of Table~\ref{Table:evolution}, the median is smaller than the mean, is consistent with the presumptions that most changes are evolutionary, and that evolutionary changes typically incur smaller size changes. However, this raw data does not provide sufficient grounds for the correct placement of any given pair between evolutionary and revolutionary extremes.

The next group of rows in \Tab{evolution} shows the statistics of the
  ratio of types (resp.\ edges) that are common to both versions of a pair,
  compared to the total number of types in earlier version (\emph{remaining})
  and the later version (\emph{continuing}).
We have that the mean fraction of remaining types is $91\%$,
    while the median fraction is $97\%$.
The fact, recurring across the entire group,
    that the median is greater than the mean,
    is, again, consistent with the presumptions that most of the
    pairs represent a more evolutionary change, in which
    most of the types remain in the subsequent version.
The fewer pairs which represent revolutionary changes
    bring the mean lower than the median.

In concentrating on the median, we see that typically
  about 3\% of types and 6\% of edges
  are lost with the release of a new version,
  and about 80\% of the types and edges in the version existed
  in the previous one.

Note that it could be the case that some of the types which were marked as removed by our analysis tools,
  were simply renamed. The extent of this analysis error is bounded above by the (small) number of removed types.

Finally, the last rows of
  the table summarize the percentage of the classes for which none of the incoming
  (outgoing) were changed during the evolution process.
We learn that about 17\% of the relations to- or from- types stay unchanged.

We can say that the \emph{functionality} of one in six types does not change, at least in the sense that
  the set of other types it uses does not change.
Also, one in six types does not change its \emph{duty} in two subsequent versions of an artifact, at least
  as far this duty is judged based on the set of other types it serves.
Conversely, \Pla{locality-of-change} is not confirmed by these results, changes to software typically border
  with 5 out of 6 types.

To summarize, the typical topological change between two subsequent versions of a software artifact
  is characterized by:
\begin{enumerate}
  \item a preservation of almost all types (3\% are lost);
  \item a preservation of almost all edges (6\% are lost);
  \item a preservation of the locale of about one sixth of the types;
  \item an increase of about 10\% in the number of types; and,
  \item an increase of about 20\% in the number of edges.
\end{enumerate}

\section{Size Changes of Artifacts in the Corpus}
\label{Section:size}
\subsection{Correlating Magnitude Changes with Version Number Changes}
Our study of the correlation between magnitude changes and version number changes,
    begins with the introduction of a notion of \emph{version number change cardinality},
    which is assigned to each pair of artifact versions by comparing the version numbers (as assigned by the artifact numbering scheme) of the pair members:
a cardinality of~$1/2^{n-1}$ is associated with
    a change to a~$n^{th}$ level version number.
Thus, the cardinality of a change to the major version number
    is~1; a change to the second level version number,
  (e.g., versions \texttt{1.3} and \texttt{1.4}) has cardinality~$1/2$, etc.

Our ensemble comprised~$\DeweyFirst$ pairs of change cardinality~$1$,%
  ~$\DeweySecond$ pairs of cardinality~$1/2$,%
  ~$\DeweyThird$ pairs of cardinality~$1/4$,%
  ~$\DeweyFourth$ pairs of cardinality~$1/8$, and~$1$ pair of cardinality~$1/16$.

\begin{figure} [!htb]
    \includegraphics[width=.5\columnwidth,angle=-90]{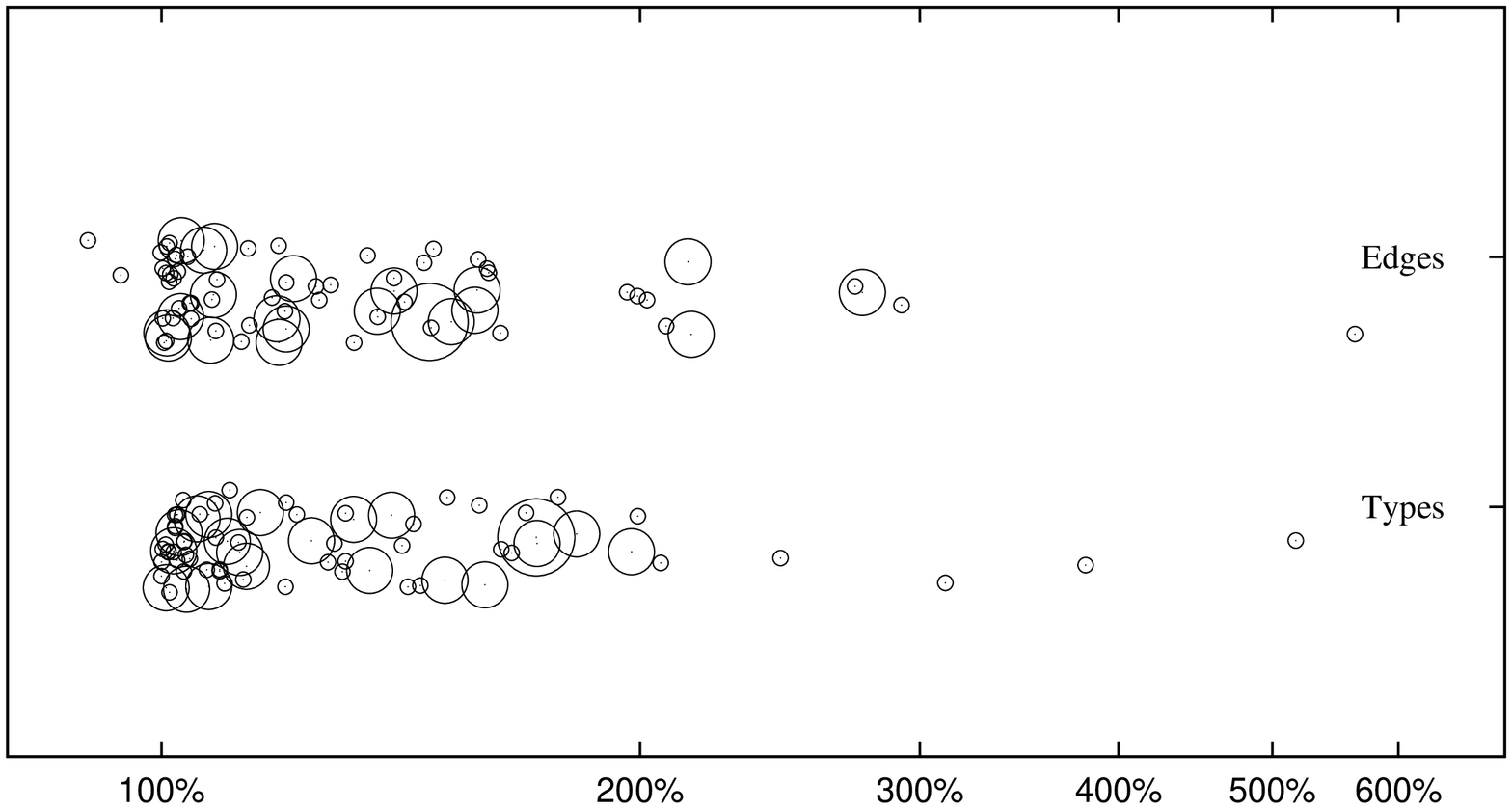}

{\scriptsize Larger circles denote more cardinal version changes}

    \caption{Distribution of relative changes in the number of edges and the number of types.}
    \label{Figure:size-change}
\end{figure}

\Fig{size-change}
  now depicts the distribution of relative changes in the number of edges
    and types in the corpus' pairs.
Each circle in the figure corresponds to a pair in the corpus;
    larger circles corresponding to more cardinal changes.

We do not know whether the more modest changes are evolutionary, that is whether these
  changes tend to preserve existing properties.
However, the picture
  depicted in \Fig{size-change} is at least consistent with \Pla{evolutionary-revolutionary-spectrum} and \Pla{mostly-evolutionary}:
  in most pairs, the increase in the number of types (edges)
  is modest; notwithstanding, a non-vanishing number of the pairs exhibit substantial increases to the number of types (edges).

Are the more drastic changes linked to the publication of major new editions of the software?
It is difficult to confirm or refute \Pla{major-changes-large} by visual inspection of \Fig{size-change},
  in trying to determine whether the larger circles tend to show on the left or on the right of the figure.
Instead, we shall describe an analytical method for studying this
    correlation using \emph{Kendall's tau coefficient}~\cite{Kendall1990}.

\subsection{Kendall's Tau Coefficient and Statistical Test}
\label{Section:Kendall}
Kendall's tau correlation coefficient gives a measure of
    the agreement between two different rankers of the same data set.
Given is a set of elements, and their relative ranking by two rankers.
Then, the coefficient is defined based on two values,~$n_c$, and~$n_d$, where~$n_c$
  is  the number of \emph{concordant}
  pairs, i.e., pairs of elements whose relative ranking by the two rankers
  are ordered in the same way, e.g., both rankers agree that the first element is ``better'' than the other,
  and where~$n_d$ is defined similarly as the number of \emph{discordant} pairs.
The simplest definition of the coefficient is~$n_c - n_d$
  divided by the total number of pairs (that is~$\binom{n}2$, where~$n$ is the number of elements in the ranked set).

The coefficient can be used for measuring the agreement of
  the ranking of a certain metric in two versions of the software.
  It assumes its maximal value of 1 in the case of full agreement,
  and its minimal value of -1 is achieved in the case of total disagreement.
The set for comparison is that of the types which occur in both versions.

Thus, Kendall's coefficient is similar to Pearson correlation, except that it is non-parametric,
  rendering it applicable to our ``change cardinality'' ranking (which is ordered, but has no obvious, non-arbitrary mapping to numerical values)
  with magnitude of change.

Herein, we used a version of the coefficient denoted~$\tau_b$ which deals with ties
  (e.g., two version pairs with the same cardinality of change).
It is computed as
\begin{equation}
\label{Equation:tau}
  \tau_b=\frac{n_c-n_d}{
  \sqrt{\displaystyle\left(\binom{n}2 - \sum_{i=1}^{k'} \binom{s_i'}2\right)
             \left(\binom{n}2 -  \displaystyle \sum_{i=1}^{k''}{\binom{ s_i ''}2}\right)}},
\end{equation}
where~$n_c$ and~$n_d$ are as before,  and where~$s_1', \ldots, s_{k'}'$
  are the size of the equivalence classes in the input set
  under one ranker (application of the metric to one version of artifact)
  while~$s_1'', \ldots, s_{k''}''$
  are the size of the equivalence classes in the input set
  under the other ranker (application of the metric to another version of the artifact).

  Kendall's tau coefficient is a non-parametric test, i.e. it does not depend on the actual values but only on their relative ranks. Thus, if some experiment was to suggest that the logarithm of the metric value should be used instead of the metric value itself~\cite{GilCohenG00}, the test results would not change.

The underlying statistical test assigns a statistical confidence level to each
  value of~$\tau_b$.

\subsection{Statistical Test of \Pla{major-changes-large}}
In comparing the ranking of the pairs by the (relative) magnitude of change and the change cardinality
  we found that~$\tau_b=0.35$ (resp.~$\tau_b=0.30$) when the size of
  the increase is measured in types (edges)
  with~$p$-value~$<0.001$.

To understand why the visual inspection of \Fig{size-change}
    does not readily yield the correlation we anticipated,
    consider the following intuitive (but not entirely exact)
    interpretation of Kendall tau's coefficient.
For types, we have that $\tau=0.35$.
Then, the probability
    of a pair (of software artifacts) of being concordant is $p=(1+\tau)/2=67.5\%\approx2/3$.
In other words, if two circles of different size
    are selected at random from the lower
    part of \Fig{size-change}, then with probability 2/3 the larger
    circle will fall to the right of the smaller circle, as opposed
    to 1/2 probability when there cardinality of version change
    is uncorrelated with magnitude.
Now, the difficulty experienced in the visual inspection
    is probably explained by the difficulty of
    distinguishing between probability $2/3$ and $1/2$,
    for such pairs, a difficulty aggravated by the fact that $31\%$
    of the pairs are of circles of the same size.

Our values of $\tau_b$ were computed across all pairs, ignoring the concern that
  increment to the second level version number in one artifact may be as drastic as
  a major version number increment in another.
In restricting the comparison to versions of the same artifact, we found
  even higher values that are statistically significant, e.g.,~$\tau_b > 0.60$.
(Notwithstanding,  artifacts with small number of versions did not yield statistically
  significant values.)

\subsection{Characteristic of Change}
\label{Section:breakdown}
We now present a topological breakdown of changes
  to the software graph.
This breakdown will be used below (\Sec{mutation})
  to guide the generation of a random mutation of a
  given software version, and for examining the reliability
  of metrics against these mutations.

Fix a pair of two consecutive versions of an artifact.
Then, three kinds of types can be distinguished:
\emph{core} types are those that are present in both versions,
\emph{removed} types are those that are present in the early version
  but not in the later version, and, conversely, \emph{new} types
  are those that are present in the newer version only.
Also, changes to edges can be further characterized:
removed edges are either \emph{core/core}, \emph{core/removed}, \emph{removed/core} or \emph{removed/removed}.
Added edges are either \emph{core/core}, \emph{core/new}, \emph{new/core}, or \emph{new/new}.
Preserved edges are always of the \emph{core/core} kind.

\Tab{evolution-detailed} summarizes the statistics
  of the changes according to this breakdown.
All values in the table are obtained by first
  normalization of the absolute numbers
  and then computing the median.
Normalization of the absolute number of edges (or types)
  was with respect
  to the number of edges (or types) in the early version.

\begin{table}[!htb]
\centering
\small
\renewcommand\arraystretch{1.1}
\renewcommand\pad{\hspace{1pt}}
\begin{tabular}{/L!!!}
Edges Kind	& \multicolumn{1}{H}{{Edges (\%)}}	& \multicolumn{1}{H}{{From Types (\%)}}	& \multicolumn{1}{H}{{To Types (\%)}}	\\
\hline % Header end	\\
Core/Core (preserved)	& 94!{\pm 6}	& 95!{\pm 4}	& 64!{\pm 12}	\\
Core/Core (added)	& 4!{\pm 2}	& 9!{\pm 6}	& 8!{\pm 4}	\\
Core/Core (removed)	& 3!{\pm 2}	& 8!{\pm 5}	& 6!{\pm 3}	\\
\hline %break	\\
Core/New	& 3!{\pm 3}	& 7!{\pm 6}	& 7!{\pm 5}	\\
New/Core	& 9!{\pm 8}	& 13!{\pm 12}	& 13!{\pm 8}	\\
New/New	& 7!{\pm 7}	& 19!{\pm 17}	& 14!{\pm 13}	\\
\hline %break	\\
Core/Removed	& 0!{\pm 0}	& 0!{\pm 0}	& 0!{\pm 0}	\\
Removed/Core	& 1!{\pm 1}	& 2!{\pm 2}	& 3!{\pm 3}	\\
Removed/Removed	& 0!{\pm 0}	& 1!{\pm 1}	& 1!{\pm 1}	\\
\hline % Footer	\\

\end{tabular}
\caption{Breakdown of added and removed edges in the corpus (median values, normalized)}
\label{Table:evolution-detailed}
\end{table}

For example, the first row of the table shows that~$94\%$ of the edges between
  core types of an early were preserved when progressing to a newer version.
Those edges originated (had them as sources) in~$95\%$ of the types;
  as their targets the edges used~$64\%$ of the types in an early version of software artifact.

The mid section of \Tab{evolution-detailed}
  reveals an interesting (but not too surprising) property
  of the ``graph cut'' separating the old and the new portions of software:
the largest bulk of added edges are those that connect newly introduced
  types to core types.
Edges in the opposite direction---leading from core types to
  newly introduced types---are rare.
The second largest bulk of added edges are among
  the newly introduced types.

\section{Metrics and Their Taxonomy}
\label{Section:metrics}
There are hundreds if not thousands metrics described in the literature. We could not test them all.
However, we tried to cover a variety of metrics and give ample consideration to the most popular ones.
This section describes the \metricsCount{}
  software metrics used in our experiments,
  and proposes a taxonomy of metrics of this sort.

\subsection{Criteria for Classification of Software Metrics}

Given is~$G$, the directed graph of software system, where each node~$v$
  represents a module of this system, and an edge~$e(s,t)$ leads from a \emph{source} node~$s$
  to a \emph{target} node~$t$ if type~$s$ uses type~$t$. A \emph{metric} then
  is a function~$\mu_G$ (or just~$\mu$ if~$G$ is clear from the context)
  that assigns a value~$\mu(v)$ to each node~$v \in G$.

\paragraph{Metric nature.}

If~$\mu(v)$ depends solely
  on the topology of~$G$, we say
  that~$\mu$ is \emph{topological}.
In contrast to topological metrics stand \emph{semantical} metrics whose value
  takes into account a deeper analysis of the node contents (by e.g.,
  examining the code in this node),
  and the sort of the edges incident on it (by e.g., distinguishing between different
  kinds of dependencies among nodes).
The suite includes \distributionSemantical{} semantical metrics.

\paragraph{Metric directionality.}

The \emph{dual} of a (topological) metric~$\mu_G$ is a metric~$\mu'_G$,
  defined by~$\mu'_G(v) \equiv \mu_{G'}(v)$ where~$G'$ is the graph
  obtained from~$G$ by inverting the direction of all edges in it.
Thus, metrics~$\mu_1$ and~$\mu_2$ are duals if~$\mu_1$ computed in ~$G$ is
  the same as~$\mu_2$
  computed in~$G'$.
A metric is \emph{undirected} if it is the dual of itself;
  it is otherwise \emph{directed}.
Our metrics suite includes \distributionDirectional{}
  directional metrics and  \distributionUndirectional{} unidirectional metrics.

\paragraph{Metric scope.}

Another criterion for classification is whether~$\mu(v)$ depends
  on~$G$ in its entirety, rather than on a restricted neighborhood of~$v$.
We say that a metric is \emph{strictly local} if~$\mu(v)$
  does not change with changes to~$G$
  that preserve incoming and outgoing edges to~$v$
  (along with the identity of the nodes at the other end of these).
In other words, metric~$\mu$ is strictly local if~$\mu(v)$ depends solely
  on~$v$ and its neighbors.
Also,~$\mu$
  is  \emph{local},  if for every~$v \in G$ there is a
  set of nodes~$S \subsetneq G$, such
  that~$\mu(v)$ does not change despite arbitrary changes
  to~$G$, as long as the nodes~$S \cup \{v\}$ and the edges among
  these are intact.

For example, the widely studied Chidamber and
  Kemerer~(CK) suite~\cite{ChidamberK94} has a number of strictly local
  methods, including \emph{Number of Children} (NOC) and
  \emph{Coupling between Object Classes} (CBO), which is defined
  as the number of types whose methods may be invoked in response to call
  to the methods of a given type.
The \emph{Depth in Inheritance Tree} (DIT) metric however is local,
  but not strictly local.

Obviously, local metrics are more suited to the study of
  a single type, or a small portion of the code;
this kind of metrics is
  not expected to be telling much of the architecture.

Overall, we have \distributionLocal{} local metrics.
A subcategory of \emph{local} metrics (\distributionInternal{} metrics in our suite) is that of internal metrics;
  a metric~$\mu$ is   \emph{internal}
  if~$\mu(v)$ depends only on~$v$.
A local metric does not make sense unless it is semantical.
\emph{Weighted Methods Per Class} (WMC)~\cite{ChidamberK94},
  is an example of an internal metric.

A metric which is not local is \emph{global},
  e.g., the PageRank metric mentioned above is global.
\ignore{At the end of reading this section,  it should be apparent
  that with all global metrics we use here, it is possible to
  mutate~$G$, without touching the neighborhood of~$v$,
  in such a way that~$\mu(v)$ assumes any arbitrary value.}

\paragraph{Metric range.}

Our fourth criterion for classifying metrics is based on the type of values
  they yield;  \emph{continuous} metrics (e.g, PageRank) yield real values,
  while \emph{discrete} metrics (e.g., CBO, NOC, and DIT) yield integers, typically
  drawn from a small range, say~$o(|G|)$.
We have \distributionContinuous{} continuous metrics,
  and \distributionDiscrete{} discrete metrics.
The remaining metrics belong to a special kind of discrete metrics
  henceforth called \emph{markers}, which yield Boolean-, that is true- or false-, values.

\subsection{Metrics Used in the Experiments}
Table~\ref{Table:metric:categories}
  enumerates the metrics used in our experiments,
  classifying these according to this taxonomy.

\begin{table}[!htb]
\centering
%\scriptsize
\renewcommand\pad{\hspace{2pt}}
\begin{tabular}{lllll}
Metric	& Nature	& Directed	& Scope	& Range	\\
\hline % Header end	\\
final	& semantical	& undirectional	& internal	& Boolean	\\
abstract	& semantical	& undirectional	& internal	& Boolean	\\
interface	& semantical	& undirectional	& internal	& Boolean	\\
\hline %break	\\
sink	& topological	& directional	& local	& Boolean	\\
source	& topological	& directional	& local	& Boolean	\\
baloon	& topological	& directional	& local	& Boolean	\\
wrapper	& topological	& directional	& local	& Boolean	\\
\hline %break	\\
pure	& semantical	& undirectional	& internal	& Boolean	\\
pool	& semantical	& undirectional	& internal	& Boolean	\\
designator	& semantical	& undirectional	& internal	& Boolean	\\
function pointer	& semantical	& undirectional	& internal	& Boolean	\\
stateless	& semantical	& undirectional	& internal	& Boolean	\\
sampler	& semantical	& undirectional	& internal	& Boolean	\\
canopy	& semantical	& undirectional	& internal	& Boolean	\\
\hline %break	\\
DIT	& semantical	& undirectional	& local	& discrete	\\
NOA	& semantical	& undirectional	& local	& discrete	\\
NOC	& semantical	& undirectional	& local	& discrete	\\
CBO	& semantical	& undirectional	& local	& discrete	\\
RFC	& semantical	& undirectional	& local	& discrete	\\
WMC	& semantical	& undirectional	& local	& discrete	\\
\hline %break	\\
\#Incoming	& topological	& directional	& local	& discrete	\\
\#Clients	& topological	& directional	& global	& discrete	\\
\#Outgoing	& topological	& directional	& local	& discrete	\\
\#Descendants	& topological	& directional	& global	& discrete	\\
\hline %break	\\
\#SCCIncoming	& topological	& directional	& global	& discrete	\\
\#SCCClients	& topological	& directional	& global	& discrete	\\
\#SCCOutgoing	& topological	& directional	& global	& discrete	\\
\#SCCDescendants	& topological	& directional	& global	& discrete	\\
SCCSize	& topological	& undirectional	& global	& discrete	\\
\hline %break	\\
\#DominatedBy	& topological	& directional	& global	& discrete	\\
\#DominatorHeight	& topological	& directional	& global	& discrete	\\
\#DominatorWeight	& topological	& directional	& global	& discrete	\\
\hline %break	\\
PageRank	& topological	& directional	& global	& continuous	\\
Betweeness	& topological	& directional	& global	& continuous	\\
Belonging	& semantical	& undirectional	& local	& continuous	\\
\hline % Footer	\\

\end{tabular}
\caption{Metrics used in experiments and their categories.}
\label{Table:metric:categories}
\end{table}

\vspace{1ex}
\noindent
\textbf{Marker Metrics.}

The first fourteen metrics in the table are markers:
  \emph{final}, \emph{abstract} and \emph{interface} are
  simply the \textsc{Java} class attributes with the same name.

Next comes a group of four topological metrics.
The \emph{sink} marker is assigned to types
  from which a bottom-up study of a software system
  may start since they are referred by
  any other type in the system (either directly or indirectly).
Conversely, the \emph{source} marker
  is for types from which a top-down study
  may start.
The \emph{balloon} marker
  (so named after balloon types~\cite{Almeida97balloontypes})
  is for types which have only one client, i.e., nodes
  whose in-degree is~$1$.
And, the \emph{wrapper} marker is just the opposite---nodes whose
  out-degree is~$1$.

Following that, we have a group of micro-patterns
  markers~\cite{Gil:Maman:05}.
For this work, we carried out measurements on seven of these.

\paragraph{Chidamber and Kemerer Metrics.}

The next six metrics are all semantical,
  undirected, discrete, and local; they were
  all drawn from Chidamber and
  Kemerer's suite~\cite{ChidamberK94}, including the metrics described
  above, together with \emph{Response For a Class} (RFC),
  which is the number of methods that can potentially
  be executed in response to an invocation of a method in the type.

The WMC metric was computed by using the
  total number of instructions in this method as method complexity.
In addition to these basic metrics, we included a variant of DIT,
  \emph{Number of Ancestors} (NOA) which seems appropriate for
  the inheritance structure of interfaces and classes in \textsc{Java}.
Of this suite~\cite{ChidamberK94}, the \emph{Lack of Cohesion} (LOC) metric was not included
  in our study.

\paragraph{Plain Topological Metrics.}
The next local metric is \emph{\#Incoming}, which counts the number of
  immediate clients a type has.
(Of  course, this metric is related to \emph{sink} and \emph{wrapper} metrics.)
In contrast,  \emph{\#Clients} is a global metric defined
  as the total number of clients of a type, including both immediate and non-immediate clients.

\emph{\#Outgoing} and \emph{\#Descendants} are the dual of
  these two, counting the number
  of types that a given type uses directly and indirectly; observe that  \emph{\#Descendants}
  is identical to Page-Jones and Constantine's~\cite[Chap.~9]{Page-Jones:Constantine:00}
  \emph{encumbrance} metric, which, according to the first author
  of this book, is indicative of the ``sophistication'' of a type, its role and
  may even be predictive of its fate.

\paragraph{Strongly Connected Components Metrics.}

The next group of metrics is computed
from the directed acyclic graph of
  \emph{strongly connected component}s of~$G$.
Recall that there is a directed path between any two nodes that reside in the same
  strongly connected component;
  this theoretical structure of a graph makes sense in a software context since
  all types in such a component are interdependent, and hence should probably be
  studied together.
A strongly connected component thus may be thought together of as \emph{super module}.
In our suite, \emph{SCCSize} represents the size
  of this super module (i.e., the size of the strongly connected component)
  that a type belongs to.
  \emph{\#SCCIncoming} and  \emph{\#SCCClients} are, respectively, the number of super-modules immediate and indirect
  clients that the super module serves. Their duals are \emph{\#SCCOutgoing} and \emph{\#SCCDescendants}.

\paragraph{Dominators Tree Metrics.}

The penultimate metrics group is computed from the
  \emph{dominators tree} of~$G$.
Recall that a node~$r$ dominates a node~$v$, if the only
  way of getting from into~$v$ is through~$r$, and that
  there is an edge in this tree if~$r$ is the ``most immediate'' dominator
  of~$v$.
Thus, the dominators tree is likely to identify pivotal points of the software
  system.
From this tree we compute the \emph{\#DominatedBy} metric which
  is the number of nodes that dominate this node, the  \emph{\#DominatorHeight},
  which is the height of the node in the dominators tree, and  \emph{\#DominatorWeight},
  giving the number of nodes that a given node dominates.

\paragraph{Other Metrics.}

In the last group of metrics in Table~\ref{Table:metric:categories} we have
  \emph{PageRank} and \emph{Betweenness}, yet another measure of graph centrality~\cite{Brandes};
  roughly speaking,  nodes that occur on many shortest paths
  connecting other nodes have higher Betweenness value than those that do not.

The last metric in the table is \emph{Belonging} used, e.g., in JDepend\footnote{\texttt{http://clarkware.com/software/JDepend.html}} and
  in SA4j\footnote{\texttt{http://www.alphaworks.ibm.com/tech/sa4j}}, which estimates
  the extent by which a type belongs to its package by dividing the number of edges
  it has (both incoming and outgoing) to other types in the package by the total
  number of edges incident on the type.

\section{Marker Metrics}
  \label{Section:marker}
  
\Tab{prevalence-marker}
  gives the essential statistics of the prevalence of the marker metrics in the suite.

\begin{table}[!htb]
\centering
\small
\renewcommand\arraystretch{1.1}
\renewcommand\pad{\hspace{1pt}}
\begin{tabular}{L!!RR}
Metric	& \multicolumn{1}{H}{{Mean (\%)}}	& \multicolumn{1}{H}{{Median (\%)}}	& \multicolumn{1}{H}{{Min (\%)}}	& \multicolumn{1}{H}{{Max (\%)}}	\\
\hline % Header end	\\
final	& 15.0!\pm 15.1	& 7.3!\pm 6.9	& 0.0	& 48.5	\\
abstract	& 4.6!\pm 2.4	& 4.1!\pm 1.8	& 0.8	& 11.8	\\
interface	& 10.1!\pm 5.2	& 8.1!\pm 4.2	& 1.7	& 21.2	\\
\hline %break	\\
sink	& 1.3!\pm 2.2	& 0.6!\pm 0.6	& 0.0	& 16.7	\\
source	& 27.7!\pm 16.4	& 29.5!\pm 13.7	& 1.0	& 55.3	\\
balloon	& 10.8!\pm 8.9	& 7.2!\pm 4.6	& 1.1	& 42.1	\\
wrapper	& 25.7!\pm 7.1	& 23.9!\pm 3.0	& 12.0	& 49.5	\\
\hline %break	\\
pure	& 8.9!\pm 5.0	& 8.1!\pm 3.7	& 0.8	& 21.2	\\
pool	& 1.4!\pm 1.2	& 1.0!\pm 0.6	& 0.0	& 5.9	\\
designator	& 0.4!\pm 0.6	& 0.2!\pm 0.2	& 0.0	& 4.3	\\
function pointer	& 0.2!\pm 0.4	& 0.0!\pm 0.0	& 0.0	& 2.2	\\
stateless	& 28.7!\pm 8.8	& 29.3!\pm 4.9	& 9.8	& 53.0	\\
sampler	& 0.9!\pm 0.7	& 0.8!\pm 0.3	& 0.0	& 3.2	\\
canopy	& 17.1!\pm 9.1	& 15.9!\pm 7.8	& 3.4	& 47.6	\\
\hline % Footer	\\

\end{tabular}
\caption{Essential statistics of the prevalence of the marker metrics in the suite.}
\label{Table:prevalence-marker}
\end{table}

  The first group of markers in the table are \textsc{Java} type attributes.
As we see, about 4\% of all types are \emph{abstract},
  about 10\% are interfaces, and 15\% are  \emph{final}.
The variance of the \emph{final} attribute
  is greater than that of \emph{abstract},
   with some versions not using it at all, while others using it for almost half of the classes.
Later we will see that this simple architecture discovery hints are surprisingly
  reliable.

The next group in Table~\ref{Table:prevalence-marker} is of topological markers.
The prevalence of sinks is low, but still could be explanatory of the architecture
  of the underlying software.
About one in three types is a source in our corpus.
This is explained by the large number of frameworks and libraries in our data set.
The resilience of these two markers to the teeth of time tends to be high,
  as we shall see shortly.

Types with only one client (balloons) are quite popular (10\%),
  but much more popular are wrapper types, i.e., those types make use of only one
 type in the artifact.

Micro patterns are the third group in the table.
Of these, it is surprising to see that almost one third of the types in the corpus are \emph{stateless}.
Even though $10\%$ of the types in the corpus are interfaces,  about $20\%$ of real classes are stateless, i.e.,
do not manage state variables at all.

Overall, we see that there is a substantial variety in prevalence of each of metrics.
  The prevalence of pure types, for example, ranges between 0.8\% and 21.2\%.
Even the smallest standard deviation, 0.4\% for the
  function pointer micro pattern, is large compared to
  its 0.2\% average prevalence.

\subsection{Reliability}

\begin{figure} [!htb]
    \includegraphics[width=0.7\columnwidth,angle=-90]{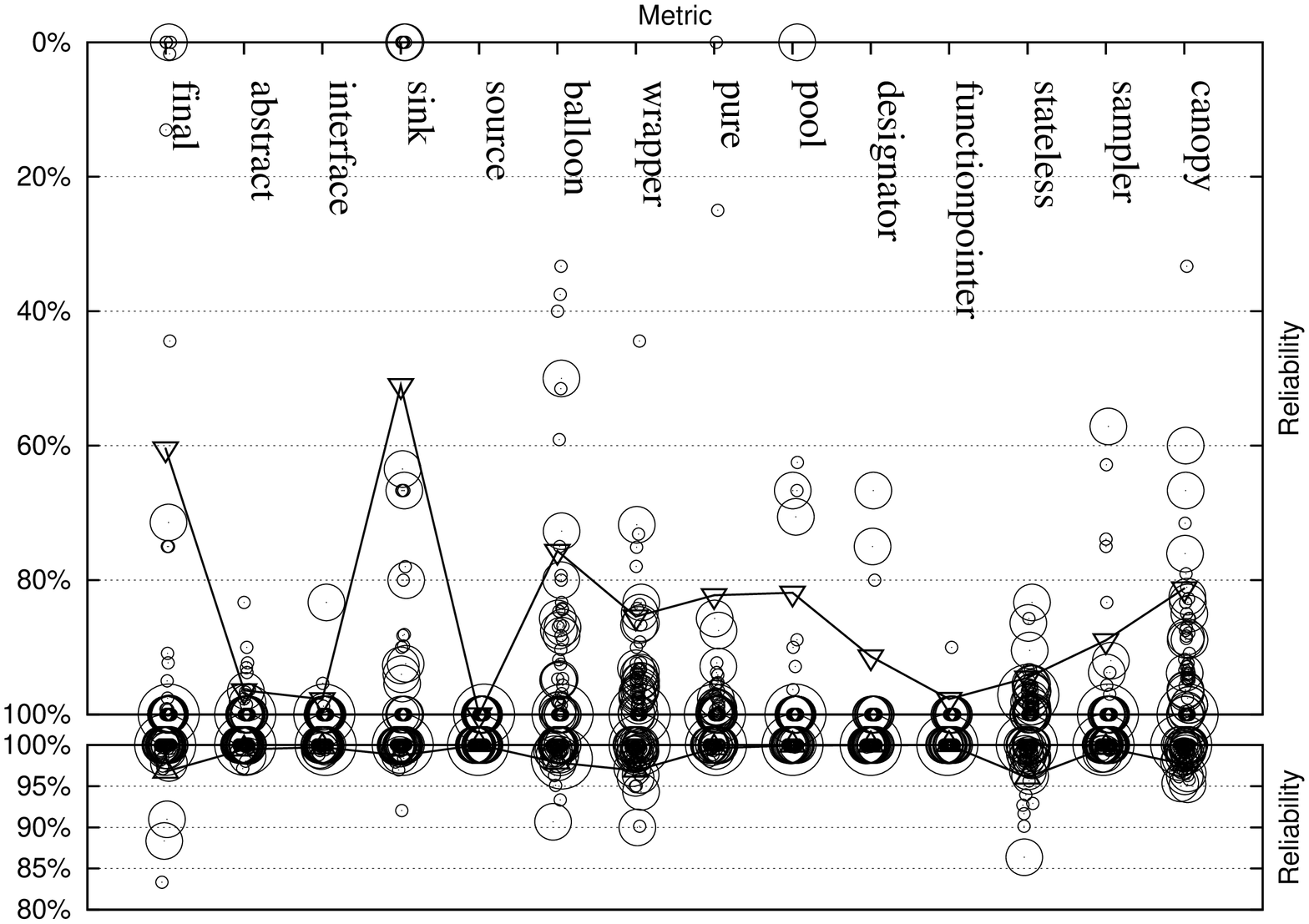}
{\small Larger circles denote more cardinal version changes; triangles denote the mean minus standard deviation barrier.}
    \caption{Reliability of marker metrics (top) and their negation (bottom) in the corpus.}
    \label{Figure:marker-reliability}
\end{figure}

The top part of \Fig{marker-reliability} depicts the reliability values of marker metrics in the corpus:
columns correspond to the metrics, while
  each circle on a column corresponds to a certain pair of consecutive versions.
The circle height represents the reliability of the marker metric in this
  pair, that is, the portion of types that preserve this marker as the software evolves from the
  earlier to the later version.

We see that reliability is generally high;
In considering, for example, the \texttt{final} marker,
  we observe that \I~in the vast majority of pairs, fewer than 5\% of
  all  \texttt{final} classes lose this property as the software evolves,
  but still, \II~there are pairs in which the loss of the \texttt{final} property occurs in
  30\%, 50\% and even 100\% of the cases.

More generally, we have that \I~in the majority of pairs,
  marker metrics are extremely reliable (the median reliability is always 95\% or higher,
  and is greater than 99\% for all but two metrics);
  however, \II~in a non-negligible number of pairs,
  a large portion of the types lose their marking.

The bottom part of \Fig{marker-reliability} depicts similar information as the top, except that it pertains to the negation of
  marker metrics, e.g., the first column of circles in the figure represents the relative number of classes
  that were \emph{not} \texttt{final} in the earlier version, but became \texttt{final} in the later version.
Even though negations are more reliable, phenomena \I~and \II~can still be discerned.

The numerical results support these observations: the median reliability is always 99\% or higher;
  it is around 100\% for the vast majority of metrics.
This, together with the fact that the mean reliability is 92\% or higher, and it is most often greater than 98\%
  confirm \Pla{metrics-reliable}.

Also, together, \I~and \II~confirm \Pla{mostly-evolutionary}.
Presumption
\Pla{evolutionary-revolutionary-spectrum} is confirmed by visual examination of the spectrum of reliability
  values in each of the 28 columns present on
  \Fig{marker-reliability}: there is no
  obvious dichotomic discretization of these spectra.%

One attempt of setting a border between the evolutionary and revolutionary ends is depicted in the figure:
  the triangles in each column denote the mean minus standard deviation level,
  but this discretization does not seem to be superior to others.
But, the partitioning between evolutionary and revolutionary ends
  offered by this line seems arbitrary, e.g., in the wrapper column,
  there does not seem to be a good reason to place the border at 70\% rather than
  at 83\%.

The size of circles in \Fig{marker-reliability} denotes change cardinality.
We expected larger circles to fall in the revolutionary end, but
  this is not to easy to confirm visually.
Instead, for \Pla{major-changes-large}, we computed~$\tau_b$ of the reliability values and the pair's change cardinality
  for each of the marker metrics (and their negations).
Similarly to the testing in \Sec{Kendall}, this computation was carried out for the entire ensemble, but
  also separately for the set of versions of each of the artifacts in the ensemble.

Rather surprisingly, the results revealed a \emph{positive} correlation of reliability values
  and cardinality.
In other words, we found that marker metrics tend to change less in major version increments and more
  in minor version changes.

The specifics are as follows: half of the~$\tau_b$ values for the entire ensemble
  were statistically significant ($p$-value less than~$0.05$); in all of these~$\tau_b$ was positive,
  ranging between 0.13 and 0.33.
Further, even all non-significant values (for the entire ensemble)
  were positive (with the sole exception of the \texttt{interface} marker metric
  in which~$\tau_b = -0.03$).
The same happens for~$\tau_b$ values computed within each artifact.
All statistically significant values are positive, ranging from~$\tau_b = 0.49$ and~$\tau_b=1$.
(Again, artifacts with a small number of versions did not usually have statistically significant results).

\subsection{Style Preservation}

Having studied changes to \emph{existing} types in our corpus, we now turn to checking
  whether newly added classes tend to follow the ``style'' of existing software body, and
  whether this tendency is correlated with the cardinality of the change.
The difficulty in testing  \Pla{preservation-of-style} is that the
  arsenal of standard statistical tests is good at showing that a set of values
  \emph{does not} follow a given (null-hypothesis) distribution, but usually
  falls short of showing the inverse---that the values' set indeed \emph{obeys} a given
  distribution.

We applied the standard~$\chi^2$ test for each of the marker metrics and each of the pairs
  in the corpus (total of~$\BatchConsecutivepairsTotal\times\distributionBoolean$ tests) to determine
  whether there is any statistical difference in the prevalence of the
  metric in the earlier software version and its prevalence among newly added types.

Statistically significant values of the $\chi^2$ value were found in only~$25.5\%$ of the tests, and,
  in confirmation of \Pla{preservation-of-style}, these significant changes to the prevalence are
  correlated with the cardinality of the change.
This correlation is not so strong,~$\tau = 0.09$, but it is statistically significant ($p$-value~$<0.01$).

Does the addition of new classes change the \emph{overall} prevalence of any of the metrics?
Our finding indicate that this rarely happens.
\Tab{flags:diff} depicts the main statistics of
  \emph{changes} to the prevalence in the corpus.

\begin{table}[!htb]
\small
\renewcommand\arraystretch{1.1}
\renewcommand\pad{\hspace{1pt}}
\centering
\begin{tabular}{L!!d2d2}
Metric	& \multicolumn{1}{H}{{Mean (\%)}}	& \multicolumn{1}{H}{{Median (\%)}}	& \multicolumn{1}{H}{{Min (\%)}}	& \multicolumn{1}{H}{{Max (\%)}}	\\
\hline % Header end	\\
final	& 0.01!{\pm 10.39}	& 0.00!{\pm 0.93}	& -46.95	& 48.54	\\
abstract	& 0.02!{\pm 1.36}	& -0.03!{\pm 0.19}	& -6.48	& 5.60	\\
interface	& -0.01!{\pm 2.23}	& -0.05!{\pm 0.50}	& -10.63	& 7.70	\\
\hline %break	\\
sink	& -0.21!{\pm 1.84}	& 0.00!{\pm 0.13}	& -9.71	& 5.44	\\
source	& 1.57!{\pm 7.13}	& 0.66!{\pm 1.04}	& -29.81	& 35.52	\\
balloon	& -0.99!{\pm 3.20}	& -0.19!{\pm 0.67}	& -20.09	& 6.80	\\
!balloon	& -0.99!{\pm 3.20}	& -0.19!{\pm 0.67}	& -20.09	& 6.80	\\
wrapper	& -0.17!{\pm 4.33}	& 0.01!{\pm 0.97}	& -18.66	& 24.49	\\
\hline %break	\\
pure	& 0.01!{\pm 2.19}	& -0.04!{\pm 0.47}	& -10.63	& 8.12	\\
pool	& 0.21!{\pm 1.00}	& -0.01!{\pm 0.12}	& -1.92	& 5.85	\\
designator	& 0.03!{\pm 0.57}	& 0.00!{\pm 0.01}	& -1.92	& 4.26	\\
function pointer	& 0.01!{\pm 0.09}	& 0.00!{\pm 0.00}	& -0.33	& 0.37	\\
stateless	& 0.69!{\pm 3.75}	& 0.10!{\pm 1.10}	& -8.13	& 15.58	\\
sampler	& -0.00!{\pm 0.51}	& -0.02!{\pm 0.08}	& -2.19	& 1.60	\\
canopy	& 0.09!{\pm 5.09}	& -0.10!{\pm 1.30}	& -13.93	& 26.65	\\
\hline % Footer	\\

\end{tabular}
\caption{Changes in prevalence of marker metrics in consecutive versions of software artifacts.}
\label{Table:flags:diff}
\end{table}

We see that in most metrics, the average change in the prevalence is about~$0.02\%$ (and is always less than~$1.6\%$).
Similarly, in most metrics the median change is about~$0.01\%$ (and is always less than~$0.7\%$).
However, it would be incorrect to say that the prevalence
  \emph{never} changes drastically in consecutive versions of software.
Examining the ``min'' and ``max'' columns
  in the table, we see that the difference
  in prevalence can be almost fifty points
  in both directions.
Even in the function pointer micro pattern, in which the extreme values
  are small, they are not negligible when compared to
  the typical occurrences of this micro pattern.
As predicted by \Pla{mostly-evolutionary}, there are few cases of revolutionary changes, in which the prevalence changes
  by as much as~$50\%$.

\section{Numerical Metrics}
  \label{Section:results}
  Reliability of marker metrics was defined simply as
    the portion of classes that retained the
    marker during the change.

A straightforward extension of this definition to
numerical metrics leads to bogus results,
  since even an addition of a single type may change the metric value of all types.
The reason is that the precise values of numerical
    metrics, even local ones, are highly sensitive to change.
For example, introducing a new root to
    the inheritance hierarchy will change the DIT metric of all types.
Our empirical findings showed that
    progressing to the next software version changed the PageRank of $99\%$ or more of the types,
    and the WMC value of the $36\%$ of the types.

We therefore use a more robust definition
    of reliability  which relies on  Kendall's~$\tau_b$ (see \Sec{Kendall} above)
    coefficient of correlation to compare the relative ordering of values that a metric yields.
The computation of~$\tau_b$ is (naturally)
    done only for the types which are present in both versions.
However, the value of~$\mu_{G_i}$
    may depend on types which only occur in~$G_i$.

High values of~$\tau_b$ imply
    high reliability.
For example, if~$\tau_b = 0.9$ for
    a certain numerical metric~$\mu$
    and a certain pair of versions~$\langle G, G^*\rangle$,
    then the implication \[
        \mu_{G}(u) > \mu_{G}(v) \Rightarrow \mu_{G^*}(u) > \mu_{G^*}(v)
  \] holds for~$95\%$
  of types~$u,v \in G \cap G^*$.

We thus computed the values of $\tau_b$ for all metrics.
This computation was restricted to consecutive pairs only for
    two reasons:
first, the reliability value in moving from version $i$
    to version $i+2$ can be broken down to, at least mentally,
    to two factors: that of the progression from version $i$
    to version $i+1$ and that of the progression from
    version $i+1$ to version $i+2$.
Second, the consideration of all pairs biases our sample
    towards artifacts with more versions.

As it turns out, the number of types involved made
    in computing~$\tau_b$, was so large
    (recall Table~\ref{Table:corpus})
    that all values thus
    computed were statistically significant with
  high confidence level ($p$-value $<0.001$.

\Tab{local:all} presents the essential statistics of these
    values for \emph{local} numerical metrics.

\begin{table}[!htb]
\small
\renewcommand\arraystretch{1.1}
\renewcommand\pad{\hspace{1pt}}
\centering
\begin{tabular}{/L!!d2d2}
Metric	& \multicolumn{1}{H}{{Mean}}	& \multicolumn{1}{H}{{Median}}	& \multicolumn{1}{H}{{Min}}	& \multicolumn{1}{H}{{Max}}	\\
\hline % Header end	\\
DIT	& 0.95!\pm 0.08	& 0.98!\pm 0.02	& 0.55	& 1.00	\\
NOA	& 0.96!\pm 0.06	& 0.98!\pm 0.02	& 0.70	& 1.00	\\
NOC	& 0.94!\pm 0.09	& 0.97!\pm 0.03	& 0.34	& 1.00	\\
CBO	& 0.93!\pm 0.07	& 0.95!\pm 0.04	& 0.59	& 1.00	\\
RFC	& 0.93!\pm 0.07	& 0.94!\pm 0.04	& 0.58	& 1.00	\\
WMC	& 0.93!\pm 0.07	& 0.94!\pm 0.04	& 0.56	& 1.00	\\
\hline %break	\\
\#Incoming	& 0.94!\pm 0.08	& 0.96!\pm 0.04	& 0.52	& 1.00	\\
\#Outgoing	& 0.93!\pm 0.06	& 0.94!\pm 0.04	& 0.67	& 1.00	\\
Belonging	& 0.87!\pm 0.13	& 0.88!\pm 0.07	& 0.22	& 1.00	\\
\hline % Footer	\\

\end{tabular}
\caption{Essential statistics of~$\tau_b$ of local metrics across consecutive versions of software artifacts.}
\label{Table:local:all}
\end{table}

Most obviously, all values are high.
In fact we have that the average value of~$\tau_b$,
    computed across all metrics and all pairs,
    is $0.93$, in confirmation of \Pla{metrics-reliable}.
We also see that the median is greater than the mean,
   confirming \Pla{evolutionary-revolutionary-spectrum}
   and \Pla{mostly-evolutionary}.

Similar finding are exhibited by \Tab{global:all} which
  repeats \Tab{local:all} for \emph{global} metrics.
Interestingly, the mean and median values of all metrics are quite similar.
Still, in comparing Mean, Median, and Min columns in the two tables
  we see that global metrics are (slightly) less reliable than local metrics.

\begin{table}[!htb]
\small
\renewcommand\arraystretch{1.1}
\renewcommand\pad{\hspace{1pt}}
\centering
\begin{tabular}{/L!!d2d2}
Metric	& \multicolumn{1}{H}{{Mean}}	& \multicolumn{1}{H}{{Median}}	& \multicolumn{1}{H}{{Min}}	& \multicolumn{1}{H}{{Max}}	\\
\hline % Header end	\\
\#Clients	& 0.93!\pm 0.09	& 0.96!\pm 0.04	& 0.55	& 1.00	\\
\#Descendants	& 0.90!\pm 0.10	& 0.93!\pm 0.07	& 0.63	& 1.00	\\
\hline %break	\\
\#SCCIncoming	& 0.90!\pm 0.12	& 0.93!\pm 0.06	& 0.48	& 1.00	\\
\#SCCClients	& 0.92!\pm 0.10	& 0.94!\pm 0.05	& 0.52	& 1.00	\\
\#SCCOutgoing	& 0.90!\pm 0.12	& 0.93!\pm 0.06	& 0.48	& 1.00	\\
\#SCCDescendants	& 0.90!\pm 0.10	& 0.93!\pm 0.07	& 0.64	& 1.00	\\
SCCSize	& 0.89!\pm 0.12	& 0.93!\pm 0.07	& 0.59	& 1.00	\\
\hline %break	\\
\#DominatedBy	& 0.87!\pm 0.17	& 0.89!\pm 0.09	& 0.05	& 1.00	\\
\#DominatedBy'	& 0.88!\pm 0.15	& 0.93!\pm 0.06	& 0.21	& 1.00	\\
\#DominatorHeight	& 0.87!\pm 0.13	& 0.90!\pm 0.08	& 0.33	& 1.00	\\
\#DominatorHeight'	& 0.88!\pm 0.15	& 0.91!\pm 0.07	& 0.10	& 1.00	\\
\#DominatorWeight	& 0.87!\pm 0.13	& 0.89!\pm 0.08	& 0.35	& 1.00	\\
\#DominatorWeight'	& 0.88!\pm 0.15	& 0.91!\pm 0.07	& 0.08	& 1.00	\\
\hline %break	\\
PageRank	& 0.93!\pm 0.08	& 0.94!\pm 0.05	& 0.50	& 1.00	\\
PageRank'	& 0.88!\pm 0.10	& 0.89!\pm 0.07	& 0.56	& 1.00	\\
Betweeness	& 0.89!\pm 0.11	& 0.91!\pm 0.07	& 0.41	& 1.00	\\
Betweeness'	& 0.88!\pm 0.12	& 0.91!\pm 0.08	& 0.35	& 1.00	\\
\hline % Footer	\\

\end{tabular}
\caption{Essential statistics of~$\tau_b$ of global metrics across consecutive versions of software artifacts.}
\label{Table:global:all}
\end{table}

\begin{figure*} [!htb]
\centering
    \includegraphics[width=0.5\columnwidth,angle=-90]{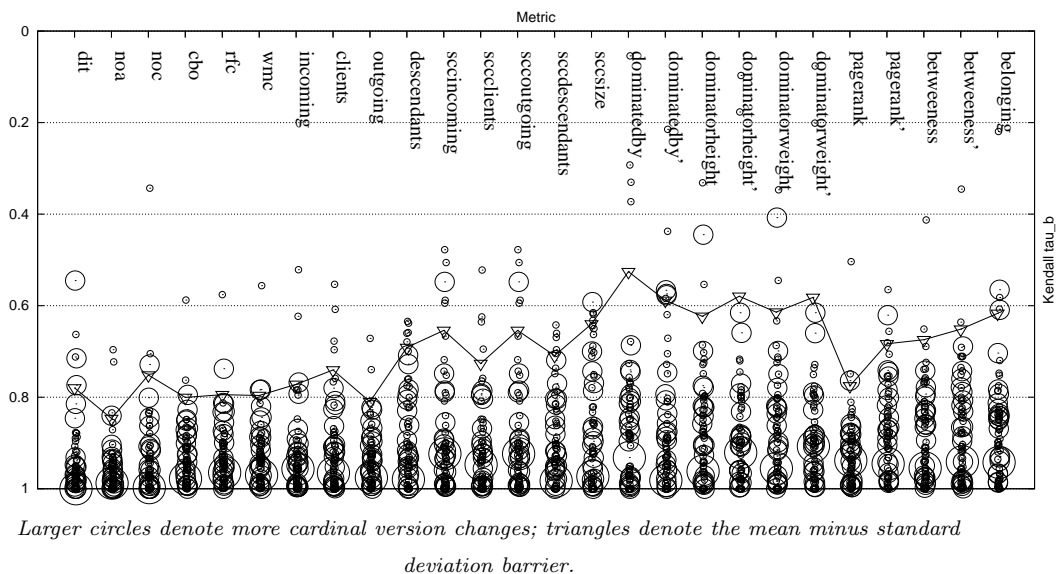}

{\scriptsize\it Larger circles denote more cardinal version changes; triangles denote the mean minus standard deviation barrier.}

    \caption{Reliability of numerical metrics in the corpus}
    \label{Figure:numeric-reliability}
\end{figure*}

\Fig{numeric-reliability} is the equivalent of \Fig{marker-reliability}
  for numerical metrics.
Examining the figure, we see again the spectrum of software changes (\Pla{evolutionary-revolutionary-spectrum}) with many
  more changes at the evolutionary end (\Pla{mostly-evolutionary}), at which reliability is high (\Pla{metrics-reliable}).
Again, a visual inspection is not sufficient to confirm \Pla{revolutionary-changes-in-major-version}, i.e.,
  that the more radical changes of the ranking offered by the metric tend to occur
  with more cardinal version number increments.

Thus, as before, to confirm \Pla{revolutionary-changes-in-major-version}, we computed for each of the metrics the coefficient of correlation between
  the ranking defined by the change cardinality number and the metric's reliability value.
This was  done for the entire ensemble.
As expected (and in contrast of what we found with marker metrics),
  the correlation was negative: higher reliability of numerical metrics tends to coincide with
  more minor increments of the version number.
Specifically, all $\tau_b$ values for the ensemble were significant and negative,
  and all values for a specific artifact which were significant were also negative.
(Values of $\tau_b$ ranged between $-0.30$ and $-0.84$, but were typically about $-0.70$.)

%\section{Reliability and Metric Scope}
%\label{Section:scope}
%\input{scope}

\section{Understanding Reliability of Global Numerical Metrics}
  \label{Section:mutation}
  How can the high reliability values of global numerical metrics be explained?
The answer that we would like to have is: ``\emph{these metrics indeed capture the essence of a software architecture;
  their preservation indicates that architecture is persistent}''.

Unfortunately, as it will become clear at the end of this section,
  this answer is only partially correct.
Much of the high agreement of the ranking is explained by the limited
  scope of changes to software between versions.
Even  \emph{random mutations}
  of the software graph reach the same high values.
In other words, most of the reliability of numerical metrics does not capture ``architecture'' in any deep way,
  and is in essence a reflection of the fact that there are many types and edges that continue
  to the next edition of a software artifact.
Interestingly, we will see that there exists a residual reliability which is well
  explained by the proviso that the relative rankings offered the global metrics suite
  capture hidden architectural traits.

\paragraph{Simple-minded, random mutation.}
Let~$G$ be a graph of a certain version of a software artifact
  and let~$G^*$ be the graph of the successive version.
Then, instead of comparing~$G$ with~$G^*$ as we did before,
  we shall compare~$G$ now with a mutation graph~$M = M(G)$, generated by random mutation of~$M$, which
  has the same number of types and edges as~$G^*$.

\Tab{implode-explode} shows the advantage of the reliability
  values found above over random, yet simple minded, graph mutations.

\begin{table}[!htb]
\centering
\small
\renewcommand\arraystretch{1.1}
\renewcommand\pad{\hspace{1pt}}
\begin{tabular}{L!!}
Metric	& \multicolumn{1}{H}{{Median (Grow)}}	& \multicolumn{1}{H}{{Median (Shrink)}}	\\
\hline % Header end	\\
\#Clients	& 0.35!{\pm 0.24}	& 0.34!{\pm 0.21}	\\
\#Descendants	& 0.46!{\pm 0.22}	& 0.46!{\pm 0.20}	\\
\hline %break	\\
\#SCCIncoming	& 0.34!{\pm 0.26}	& 0.35!{\pm 0.23}	\\
\#SCCClients	& 0.32!{\pm 0.22}	& 0.37!{\pm 0.21}	\\
\#SCCOutgoing	& 0.34!{\pm 0.26}	& 0.35!{\pm 0.23}	\\
\#SCCDescendants	& 0.41!{\pm 0.23}	& 0.40!{\pm 0.22}	\\
SCCSize	& 0.44!{\pm 0.26}	& 0.44!{\pm 0.21}	\\
\hline %break	\\
\#DominatedBy	& 0.50!{\pm 0.27}	& 0.43!{\pm 0.25}	\\
\#DominatedBy'	& 0.35!{\pm 0.27}	& 0.38!{\pm 0.24}	\\
\#DominatorHeight	& 0.45!{\pm 0.24}	& 0.36!{\pm 0.23}	\\
\#DominatorHeight'	& 0.33!{\pm 0.23}	& 0.28!{\pm 0.20}	\\
\#DominatorWeight	& 0.44!{\pm 0.24}	& 0.35!{\pm 0.23}	\\
\#DominatorWeight'	& 0.31!{\pm 0.24}	& 0.27!{\pm 0.19}	\\
\hline %break	\\
PageRank	& 0.23!{\pm 0.15}	& 0.21!{\pm 0.14}	\\
PageRank'	& 0.30!{\pm 0.15}	& 0.28!{\pm 0.15}	\\
Betweeness	& 0.39!{\pm 0.14}	& 0.35!{\pm 0.15}	\\
Betweeness'	& 0.38!{\pm 0.14}	& 0.34!{\pm 0.16}	\\
\hline % Footer	\\

\end{tabular}
\caption{Difference between reliability of global metrics across consecutive versions of software artifacts
  and reliability computed in a random graph growth and shrink.}
\label{Table:implode-explode}
\end{table}

In the experiments, we computed, for each of the metrics,
  the reliability of the metric in the pair~$\langle G,G^* \rangle$
  and subtracted from it the reliability of this metric in
  the pair~$\langle G,M_0\rangle$ where graph~$M_0$ was generated by adding
  to~$G$ random nodes and edges necessary to make it as large as (occasionally as small as)~$G^*$.
The results were then summarized in the second table column.

The third table column was computed in a similar fashion, except that
  it uses a variant method of computing the mutation~$M_0$:
  instead of growing~$G$, we randomly shrink~$G^*$
  to obtain~$M_0$, and used the pair~$\langle M_0,G\rangle$
  instead of~$\langle G,M\rangle$.

Comparing both the second and third columns of this table with the high values
  found in \Tab{global:all} shows that
  these high numbers are not so telling.
Half to two thirds of the apparent agreement
  between metric values of two versions of the software
  is found in totally random mutations.

We observe still that the agreement between metric values in a real successor is \emph{always}
  better than that of a random mutation.
Having this happen \emph{consistently} in 17 metrics cannot be a mere coincidence.
We have that with statistical significance
  of~$\alpha < 0.001$ or better, the agreement of
  metrics ranking is not a matter of pure chance.

\paragraph{Topological mutations.}
The randomness allowed in the above mutations
  allowed situations which are unlikely
  to occur in the life-cycle of software.
For example, in selecting edges in a complete random fashion,
  the number of edges between the existing nodes
  and the new nodes would be much smaller than in the real new version graph~$G^*$.
We ask now whether there exists a more structured mutation that
  can yield  the \emph{same} reliability values as found in actual software evolution.

The five mutations presented next try to imitate
  the topology of the changes to a software graph.
All of these mutations start with the original graph~$G$
  and apply two preliminary transformations to this graph:
First, all edges and nodes present in~$G$
  but not in~$G^*$ are removed.
Second, all nodes present in~$G^*$ are added.

These deterministic transformations create a graph
  which \I~has  all the core nodes,
  \II~has all the preserved core/core edges,
   and \III~has the same number of nodes as~$G^*$.
The duty of a subsequent random mutation is to add
  new edges to this transformed graph so
  that it has the same number of edges as~$G^*$.

Recall now our categorization of edge kinds and their fate
  (\Sec{breakdown}).
Edges in graph~$G^*$ are of the following five kinds:
  core/core (preserved), core/core (added), core/new, new/core and new/new.
Our construction of the transformed graph is such that the
  first kind exists in it.
All mutations in our experiments add the correct
  number of missing edges of each of the
  four remaining kinds.

The difference between the mutations is in the way the source
  and target nodes are selected for edges in each of these categories.
We use three different policies.
\begin{enumerate}
  \item
  \textsf{Same} means that the source and the target are not selected at random;
        we simply copy the edge from~$G^*$ to the mutated graph;
\item
  \textsf{Random} means that the source and the target are selected at random from the
        sets of all nodes in the corresponding group.

For example, if we apply this policy of selection to generate random core/new edges,
        then every such edge connects a random node in the core with a random
        edge in the new set.
\item
  \textsf{Random/boundary} is similar to the \textsf{Random} policy, except
        that the source and the target are selected at random from the more
        restricted set of nodes which served in~$G^*$ as source or target for the corresponding
        kind.

        For example, in applying this policy of selection to generate random core/new edges,
          every newly created edges
              connects \emph{(i)} a node selected at random from the set of core nodes
              with an edge in~$G^*$ leading to a new node with a \emph{(ii)} a node selected at
                random from the set of new nodes
              with an incoming edge in~$G^*$ starting at a core node.
\end{enumerate}

\Tab{mutations} uses these policies
  to describe the mutations we apply.
Columns of the table describe the locus of mutations:
  the \emph{\textbf{core}} locus refers to added core/core edges;
  the \emph{\textbf{cut}} locus refers to the core/new and new/core edges; and,
  the \emph{\textbf{new}} locus refers to new/new added edges.

\begin{table}[!hbt]
\centering
  \begin{tabular}{c|c|c|c}
         & \emph{\textbf{Core}} & \textbf{\emph{Cut}} & \textbf{\emph{New}} \\
\hline
 ~$M_1$  &  \textsf{Random}   & \textsf{Random}      & \textsf{Random} \\
 ~$M_2$  &  \textsf{Random/boundary}  & \textsf{Random/boundary}      & \textsf{Random/boundary} \\
 ~$M_3$  &  \textsf{Same}  & \textsf{Random/boundary}      & \textsf{Random/boundary} \\
 ~$M_4$  &  \textsf{Random/boundary}  & \textsf{Random/boundary}      & \textsf{Same} \\
 ~$M_5$  &  \textsf{Same}  & \textsf{Random/boundary}       & \textsf{Same} \\
\hline
  \end{tabular}
   \caption{Mutations imitating a subsequent software version}
   \label{Table:mutations}
\end{table}

Mutation~$M_1$ is the simplest; it is much like~$M_0$
  described
  above, except that it maintains
  the balance of edge groups.
Mutation~$M_2$ imitates slightly better a
  real software version, in that it uses the \textsf{Random/boundary} policy
  for edge selection.

Mutations~$M_3$,~$M_4$ and~$M_5$ were designed with the objective of  understanding
  better which graph locus contributes more to the agreement of
  metric values:
\begin{itemize}
\item Mutation~$M_5$ is almost identical to~$G^*$.
The difference is only in a random selection
  of edges in the \emph{\textbf{cut}} graph locus
  (and, even these edges are not entirely random; they only
  connect nodes which were adjacent on the \emph{\textbf{cut}}
  in graph~$G^*$).
\item
Mutations~$M_3$ and~$M_4$ are similar to~$M_5$ except that
  in~$M_3$ there are random changes to the core locus
  and in~$M_4$ there are random changes to the new locus.
\end{itemize}

\begin{figure*} [!htb]
\centering
    \includegraphics[width=1\textwidth]{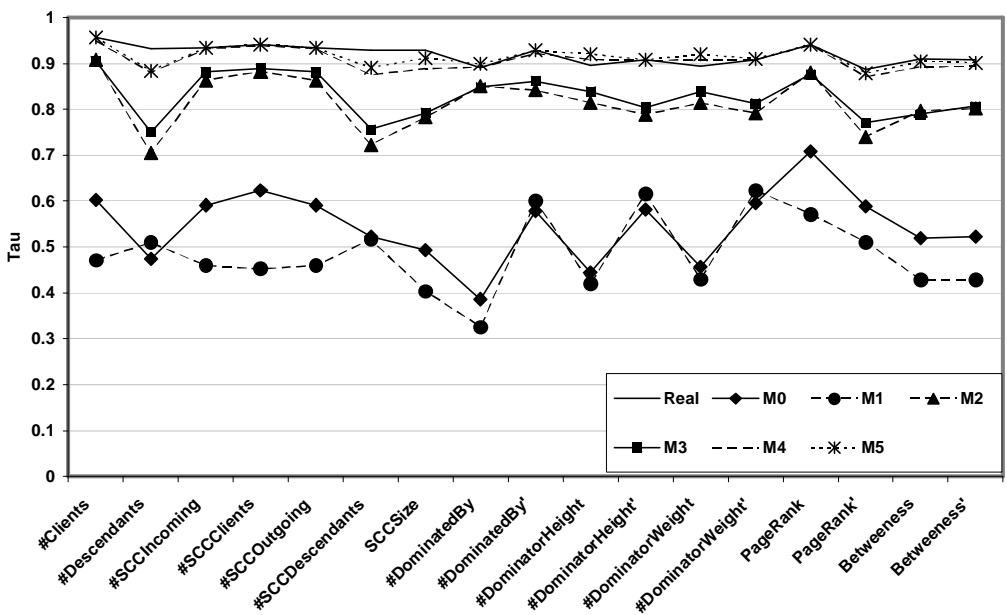}
    \caption{Median value of reliability of global metrics across consecutive versions of software artefact's and random mutations~$M_0,\ldots,M_5$.}
    \label{Figure:random-compare}
\end{figure*}

\paragraph{Results.}
\Fig{random-compare} summarizes the median of the reliability values calculated for the
  pair~$\langle G,G^* \rangle~$ and~$\langle G, M_i \rangle$
  for~$i=0,1,\ldots,5$.

The top most line in the figure, labeled ``real'' designates the high
  agreement values found in the~$\langle G,G^* \rangle~$ pairs.
The other lines correspond to the mutations.
We see that the random and not so structured
  mutations~$M_0$ and~$M_1$ explain about 2/3 of the high values
  of agreement.
Furthermore, even though~$M_1$ is more structured,
  it is inferior to~$M_0$ in at least several metrics.
A substantial improvement occurs when we move to~$M_2$
  in which we select an edge to connect two random
  ``portal'' nodes.

The reliability values of~$\langle G, M_5 \rangle$ are almost
  exactly the same as those of real software versions, i.e., of~$\langle G, G^* \rangle$.
This is not very surprising, since the graph~$M_5$ is different from~$G^*$-
  only in one locus.

Now both~$M_3$ and~$M_4$ are different from~$G^*$ in two loci.
We expect~$M_3$ to agree with~$G$ better than~$M_4$ agrees with it.
The reason is that the agreement between metric rankings is compared only at
  core nodes.
As indicated by \Tab{mutations}, the edges of~$M_3$ at the core are exactly the same as in~$G^*$.
We expected mutation~$M_3$ to yield good reliability values:
after all, in this mutation randomness was limited
  to the \emph{\textbf{new}} locus, which is the farthest from the core.

What is surprising though is that mutation~$M_4$ (in which the core/core added edges are random)
   approximates a real software version almost at the same level of agreement as~$M_3$.
Put differently, the \emph{\textbf{new}}  and the \emph{\textbf{core}} loci have the same impact on
  reliability.

\paragraph{Analysis}
We have thus observed that the particular way in which real evolution of software ``selects'' edges
  in the \emph{\textbf{new}} locus has a substantial and measurable impact on the reliability
  of global numerical metrics.
This observation is consistent with \Pla{preservation-of-style}.

More importantly, the dual of this observation tells us that our suite of
  global numerical metrics is sensitive to additions to the \emph{\textbf{new}} locus.
Only if these additions are done in a manner ``consistent'' with evolution of real software,
  reliability is preserved.

Perhaps the simplest explanation of this observation and its dual is in the
  claim that this suite reflects an underlying \emph{software architecture}.
If the introduction of edges to the \emph{\textbf{new}} locus is consistent
  with the underlying architecture, rankings of nodes as defined by the suite
  will not change much.
In contrast, random additions to the  \emph{\textbf{new}} locus, which are inconsistent with the architecture,
  will perturb these rankings.

\section{Related Work}
  \label{Section:related}

A measure of the relative importance of components within the software structure was examined in~\cite{NeateIC06}. The authors suggested to use \textsc{CodeRank}, the software equivalent of Google's well known \textsc{PageRank}~\cite{page1999pagerank} method for ranking web pages, metric to indicate how important a specific component is based on its coupling to the rest of the system. In an earlier work~\cite{Katsuro2003}, for the same purpose the authors suggested to use a similar metric called \textsc{Component Rank}. The main difference between these metrics is that the \textsc{CodeRank} is computed based on the weighted graph that represents various usage relations between the components and the number of time each usage occurs.

Lorenz et al.~\cite{Lorenz1994OOSoftwareMetricsGuide} recommend using
  a wide range of metrics to test the quality of models, classes and methods.
Various metrics related to coupling, inheritance and size of classes and methods play the major role in
  deducting the quality of the software. However, the reliability of those metrics is not analyzed by the authors.

Lajios et al.~\cite{Lajios2009MetricsSuites} investigated the correlation of
various software metrics to the defect found in software modules and proposed
an approach to determine a sets of metrics for quality assessment of complex software systems.
First they calculated various quantitative, complexities, coupling and other
metrics at the class level for several similar projects using different open source tools.
Then they found the correlation of these metrics to the history of bugs using machine learning techniques.
They found that although some of the metrics are more suitable for the assessment of software quality,
these metrics differ between the analyzed projects even though their natures are similar.
They also discovered that 5 out of 11 metrics were irrelevant for the analyzed systems.

Ordonez et al.~\cite{Mauricio2008StateOfMetrics} examined various metrics used in software industry to
measure code size and design complexity.
They mentioned that NASA used the first five metrics presented in~\cite{ChidamberK94} in the
tool they developed for analyzing source code with respect to its architecture.
The author's analysis was focused on how reliable are specific software modules
with respect to their maintainability and the probability of causing defects.
They, however, did not explore whether the metrics themselves were reliable.

\ignore{
Abdeen et al.~\cite{abdeen} present a heuristic search-based approach for optimizing inter-package dependencies of object oriented systems.
The technique minimizes the connectivity among the packages as inspired by cohesion and coupling principles.
The technique is limited to direct cyclic-connectivity and enables only reorganization of classes over existing packages.
The authors specify the modularization quality metrics and
evaluate the method against these metrics on several applications with different original modularization. The results show that a significant reduction of package coupling and cycles was achieved only by moving a relatively small number of classes.

}

Kitchenham et al.~\cite{KitchenhamS97} compared the
ways the axioms sets are derived and used in mathematics with those used in software metrics research.
\ignore{The authors classified axioms in both areas and pointed out several major differences between them.
In mathematics, for example, the main use of axioms sets is to allow logical deduction of new facts,
while in software the axioms are mostly used to delimit some class of entities to ensure the validity of new metrics.
Another major difference is that in mathematics the axioms specify conditions that are both necessary and sufficient,
while in software they give only necessary conditions. In addition software metrics axioms set is not based
on the valid example measures and there is no precise definitions of the data that they apply.
Finally, as the consistent axiom set cannot be assumed as valid, the authors suggest checking the axiom
set against the adequacy criteria to see if the things that satisfy the axiom are the things that we are trying to describe. }The authors claim that the use of axioms for measurement of size and complexity concepts is not mature enough and that there is a non-negligible risk that using axioms to validate software metrics may reject a valid measure or accept an invalid one.

The issue of validity was more frequently discussed in the community than reliability.
For many years, researchers argued (see e.g.,~\cite{Meyer1998}) that it is very difficult
to come up with a solid proof that any external metrics and measurements of software,
such as Halstead's software science metrics~\cite{Halstead1977}, and even cyclomatic complexity~\cite{McCabe76} or
even size, pertains to more interesting,  internal properties, such as maintainability,  stability, etc.
The community therefore resorts to convincing argumentation, often backed by mathematical arguments~\cite{McCabe1989},
case study analysis~\cite{Tip2008,Tip2000}, etc.~\cite{Ducasse2001blueprint}.

Mahmoud et al.~\cite{Mahmoud2003} investigated the logical stability of object-oriented designs. They computed the correlation between Chidamber and Kemerer's metrics~\cite{ChidamberK94} and the likelihood of classes to stay change-prone as a consequence of changes made to other classes in the design. They analyzed a list of design and class-level changes, and investigated how changes in one class affect others. They showed that five out of six metrics were negatively correlated with the logical stability of the classes. The authors performed their analysis on a relatively small set of subject systems and on a single version of each system. They focused on local metrics without analyzing the actual changes made on systems as they are being developed.

Fenton et al.~\cite{Fenton1999} investigated the metric based software defect prediction models and suggested that various size and complexity metrics can not serve as good predictors to software defects. They criticized the approaches that used some of the metrics covered in this work with respect to defect prediction. However, they did not question the reliability of the metrics with respect to the design of the software. Furthermore, the changes of metric values and the number of defects over time were excluded from their prediction model.

Emam et al.~\cite{Emam2001} argued that the validity of object-oriented metrics is questionable as most of them are indifferent to the size of the software. They examined Chidamber and Kemerer~\cite{ChidamberK94} metrics as well as Lorenz and Kidd~\cite{Lorenz1994OOSoftwareMetricsGuide} metrics and showed that their correlation with fault-proneness is similar to their correlation with the number of source code lines. Therefore they claimed that any analysis of metrics for software artifacts should be ``normalized'' with respect to their size. Evanco~\cite{Evanco2003} criticizes the statistical analysis proposed by Emam and claims that the model suggested by the authors fails to provide useful information as to the effect of the size of the code. In our work we covered software artifacts of various sizes and found that some metrics were reliable even when the size of the same software artifact increased five times between two consecutive versions of the same artifact.

\section{Conclusions}
  \label{Section:conclusions}

We presented a metrics suite comprising 36 code metrics drawn from various independent sources.
Our taxonomy of metrics included a distinction between semantical and topological metrics,
  a breakdown by directionality, and range of values yielded by the metric.
Our study did not reveal substantial differences between semantical and topological metrics.
Also, the distinction between directional and uni-directional metrics did not translate into
  different properties of the metric.
We did not identify any meaningful distinction between discrete and continuous metrics,
  even though some of the metrics with discrete values assumed only~$4$ or~$5$ different values.

Curiously, even though the metrics originated from different sources and described
  by different authors, all metrics in the same group had essentially the same behavior:
for example, reliability of all marker metrics was about~$99\%$,
and reliability of all (but one) local metrics was about~$93\%$.

Most of the presumptions presented in \Sec{introduction} were confirmed.
Exceptions were: \Pla{locality-of-change}, for which we found that~$5$ out of~$6$ types included at least one changed
  type in their neighborhood, and \Pla{revolutionary-changes-in-major-versions} whose opposite was
  confirmed with respect to local metrics. The presumption \Pla{metrics-reliability} was not confirmed 
  for all the analyzed metrics: marker metrics were reliable. For numerical metrics, 
  our experiments showed that reliability was
    negatively correlated with scope:
internal metrics, i.e., metrics which depend only on a certain
  class were extremely reliable;
local metrics which depend on a class and its neighbors
  were slightly less reliable;
global or topological metrics
  were unreliable.

Presumption \Pla{preservation-of-style} was confirmed under an implicit interpretation of the term ``style''
  as the prevalence of marker metrics.
In a sense, \Sec{mutation} tried to explore this presumption from the point of view of global numeric metrics.
It was shown in this section that the interconnection between newly added types have a strong impact on
  the ordering of global numeric metrics computed at the core types.

Further research should probably focus on the link between numerical metrics and 
topological architecture as implied by the phenomena shown in Section~\ref{Section:mutation}.

\balance

\bibliographystyle{abbrv}
\bibliography{00}
\end{document}